\documentclass[12pt]{article}

\usepackage{times}
\usepackage{bm,url}
\usepackage{graphicx}
\usepackage{amsfonts}
\usepackage{amsmath}
\usepackage{amssymb}
\usepackage{bm}
\usepackage{graphicx}
\usepackage{enumerate}
\usepackage{booktabs,caption,fixltx2e,subfigure}
\usepackage[flushleft]{threeparttable}

\usepackage[round]{natbib}
\usepackage{setspace,cite}
\begin{document}

\def\Diag{\textrm{diag}}
\def\inv{^{-1}}

\title{\huge RAPTT: An Exact Two-Sample Test in High Dimensions Using  Random Projections\vspace{0.in}}

\author{
 \textbf{Radhendushka Srivastava}\\
         Department of Statistical Science\\
       Cornell University\\
          Ithaca, NY 14853, USA\\
       \texttt{rsrivastava22@gmail.com}
       \and
 \textbf{Ping  Li}\\
         Department of Statistical Science\\
       Cornell University\\
          Ithaca, NY 14853, USA\\
       \texttt{pingli@stat.rutgers.edu}
       \and
 \textbf{David Ruppert}\\
         Department of Statistical Science  and  \\ School of Operations Research and Information Engineering\\
       Cornell University\\
          Ithaca, NY 14853, USA\\
       \texttt{dr24@cornell.edu}
}

\date{}

\maketitle

\vspace{0.1in}
\begin{abstract}

In\footnote{This is a revised version from a paper submitted on Feb. 02, 2013.}
high dimensions, the classical Hotelling's $T^2$ test tends to have low power or becomes undefined due to singularity of the sample covariance matrix. In this paper, this problem is overcome by projecting the data matrix onto lower dimensional subspaces through multiplication by random matrices. We propose RAPTT (\textbf{RA}ndom \textbf{P}rojection \textbf{T-T}est),  an exact test for equality of means of two normal populations based on projected lower dimensional data. RAPTT does not require any constraints on the dimension of the data or the sample size. A simulation study indicates that in high dimensions the power of this test is often greater than that of competing tests. The advantage of RAPTT is illustrated on high-dimensional gene expression data involving the discrimination of tumor and normal colon tissues.
\end{abstract}

\newpage\clearpage

\section{Introduction}\label{s1}
Conventional multivariate statistical methods are generally derived under a set-up where the data dimension ($p$) is smaller than the sample size ($n$). It is known that some of these methods either become undefined or perform poorly on a high dimensional dataset, i.e., when $p>n$.  Testing of the equality of means among high-dimensional populations occurs, for example,  in biological applications \citep{2,3}.
The limitation of conventional methods in high dimensions has led researchers to look for alternatives. For example, \citet{VB,KM,KV,9,10,1} studied inference for means in high dimensions; \citet{FHY} worked on simultaneous testing of means based on marginal tests in high dimensions; \citet{BH} worked on controlling false discovery rates in multiple testing; \citet{LW,CZZ,LC} considered testing of covariance matrices in high dimensions.

Consider testing for the equality of means of two independent multivariate normal populations.
Let  ${\bf X}_{n_1\times p}$ and ${\bf Y}_{n_2\times p}$ be data matrices with rows independently distributed as $N_p(\mu_1,\Sigma)$ and $N_p(\mu_2,\Sigma)$, respectively, where $\mu_1$ and $\mu_2$ are the respective mean vectors and $\Sigma$ is the common covariance matrix. The hypotheses are
\begin{eqnarray}\label{hyp}
{\bf H_0}:\mu_1=\mu_2 &\mbox{versus}&
{\bf H_1}: \mu_1\ne\mu_2.
\end{eqnarray}
The well-known Hotelling's $T^2$ test statistic for this testing problem is
\begin{eqnarray}\label{Tstats}
T^2= \frac{n_1n_2}{n_1+n_2} (\overline{X}-\overline{Y})' S^{-1} (\overline{X}-\overline{Y}),
\end{eqnarray}
where the sample means are $\overline{X}=\frac{1}{n_1}\bf{X}'\bf{1}$ and $\overline{Y}=\frac{1}{n_2}\bf{Y}'\bf{1}$, $\bf{1}$ is a vector of ones, $S=\frac{1}{n}\left[S_X+S_Y\right]$ is the pooled sample covariance matrix, $n=n_1+n_2-2$, $S_X={\bf{X'E_1X}}$, $S_Y={\bf{Y'E_2Y}}$, ${\bf{E_1}}={\bf{I}}_{n_1\times n_1}-\frac1{n_1}{\bf1}{\bf1}'$, ${\bf{E_2}}={\bf{I}}_{n_2\times n_2}-\frac1{n_2}{\bf1}{\bf1}'$, and $\bf{I}$ is the identity matrix.

The $T^2$ test can be used for the two-sample problem only when $p< n$ \citep{Mardia}.   When $p>n$, the pooled sample covariance matrix $S$ is singular so the $T^2$ statistic (\ref{Tstats}) becomes undefined.
Moreover, \citet{BS} showed that even when $p< n$ but the ratio $p/n\approx1$, the power of Hotelling's test is very small.  In this paper, $a \approx b$ means that $a/b\to 1$.

Several researchers have attempted to extend Hotelling's $T^2$ statistic to the $p>n$ situation by replacing $S$ with a nonsingular matrix. \citet{BS} proposed a test (referred to here as the {\it BS test}) based on the statistic $(\overline{X}-\overline{Y})'(\overline{X}-\overline{Y})$.
They established the asymptotic normality  under ${\bf H_0}$ of this statistic with appropriate location and scale transformation in the set-up where $p,n\rightarrow\infty$ such that $p/n\rightarrow c<\infty$. \citet{CQ} modified the BS test (referred to here as the {\it CQ test}) and showed that the same asymptotic power could be achieved even if $p/n\rightarrow\infty$. In another approach, \citet{SD} considered the statistic $(\overline{X}-\overline{Y})' [\textrm{diag}(S)]^{-1} (\overline{X}-\overline{Y})$ and proposed a test (referred to here as the {\it SD test}) based on asymptotic normality  under ${\bf H_0}$ of this statistic with appropriate location and scale transformation. They showed that, under certain alternatives, the asymptotic power of {\it SD test} is superior to that of {\it BS test}. In an earlier work, \citet{Sri} modified the $T^2$ statistic by replacing the inverse of $S$ with the Moore-Penrose inverse of $S$ and proposed a test based on asymptotic normality  under ${\bf H_0}$ of this modified $T^2$ statistic with appropriate location and scale transformation. In another approach, \citet{LJW} proposed an asymptotic test (referred to here as the {\it LJW test}), based on a randomized projection technique. They replaced $S$  in $T^2$ by $E_{\bf{\mathbb{R}}}\left[{\bf{\mathbb{R}}}({\bf{\mathbb{R}}}'S{\bf\mathbb{R}})^{-1} {\bf{\mathbb{R}}}'\right]$  where $\bf{\mathbb{R}}$ is random matrix of order $p\times k$ and $E_{\bf{\mathbb{R}}}[\cdot]$ is the expectation operator over the distribution of $\bf{\mathbb{R}}$. They showed that the modified $T^2$ statistic is asymptotically normal under ${\bf H_0}$ with appropriate location and scale transformation  in the set-up  where $p,n\rightarrow\infty$.

\citet{CPPW} regularized Hotelling's $T^2$ test for pathway analysis in proteomic studies by replacing $S$ with $S+\lambda I$, where $\lambda>0$. They proposed a bootstrap one sample test for high dimensional data.
\citet{WPQ} proposed a jackknife empirical likelihood test (referred to here as the {\it WPQ test}) for the equality of means in high dimensions. Under some conditions on moments, they showed that the null asymptotic distribution of the empirical likelihood is $\chi^2$ with degree of freedom 2. Here, we should point out that the asymptotic null distribution is derived under the setup $p=o\left(n^{\frac{\delta+\min(\delta,2)}{2(2+\delta)}}\right)$, where $\delta>0$, and is related the conditions on the moment.

It is important to note that the BS, CQ SD and LJW tests are asymptotic tests, and the asymptotic null distributions of the respective test statistics are derived under the set-up where $p,n\rightarrow\infty$. The bootstrap test proposed by \citet{CPPW} is also based on the asymptotic distribution of the regularized Hotelling's $T^2$. Further,  a jackknife empirical likelihood test proposed by \citet{WPQ} is also based on an asymptotic null distribution. In high-dimensional gene expression microarray applications, one often encounters a few dozen samples with dimensions in the hundreds or thousands. Asymptotic expressions may not always work well when the sample size is so small  relative to the dimension.
Moreover, the power of these asymptotic tests depends upon the structure of the covariance matrix $\Sigma$. In the absence of knowledge about this structure, it is not clear which test would generally have larger power.

In  small samples, an exact bootstrap test is usually preferred over an asymptotic test. The reasons for this preference are well-known (see \citet{MacKinnon}, and references therein).
In the present paper, we propose a randomized extension of Hotelling's test that we call RAPTT (RAndom Projections T-Test) which involves randomly projecting $p$-dimensional samples into a space of lower dimension $k$, where $k<n$.

Each random-projection Hotelling test statistics has the usual, well-known distribution, so an exact p-value can be computed.  RAPTT is based upon the average p-value over many independent random projections.  The null distribution of the average p-value does not depend on unknown parameters, only on the known distribution of the random projection matrix, and so can be computed by simulation.  Therefore, RAPTT is an exact test if we ignore Monte Carlo error, which of course can be made arbitrarily small by using a large enough Monte Carlo sample size. In addition, in the high dimensional framework where $p/n$ to tend to a positive constant or infinity, we derive the asymptotic power function of RAPTT.

RAPTT is different from past work in the
way that covariance structure is incorporated into the test statistic. The previously described test
statistics of BS, CQ and SD are essentially based on versions of the Hotelling $T^2$
test using diagonal estimators of $\Sigma$.
Our empirical study shows that this type of biased estimation of $\Sigma$ sacrifices power
when the variables are correlated, or when most of the variance can be explained by a small
number of variables in small $n$, large $p$ situations.
RAPTT achieves its power by utilizing the complete covariance matrix.
We note that the use of projection-based approaches to two-sample testing and covariance
estimation have also been considered previously by \citet{9,10,11,12}.

\vspace{0.1in}

The paper is organized as follows.  In Section~\ref{s2},  theoretical properties of the Hotelling test based on a single random projection are established.  In Section~\ref{s3}, we propose RAPTT based on the p-values of Hotelling tests from an independent sample of projections.   We discuss how critical values for RAPTT can be calculated exactly by simulation.  In Section~\ref{sR} we discuss the choice of the random projection matrices.  In Section~\ref{s4}, we present an extensive simulation study to compare the finite sample performance of RAPTT with the  asymptotic tests discussed previously. RAPTT is applied to a gene expression example in Section~\ref{s6}.  Some concluding remarks are in Section~\ref{s5}.  The proofs are provided in the Appendix.

\section{The Random-Projection Hotelling Test}\label{s2}

As already mentioned, if $p>n$ then Hotelling's test  (\ref{Tstats}) is undefined.  Our proposed solution is based on the results that the random projection of a vector can reduce its dimension and the norm of the projected vector can be made arbitrarily close to that of the original vector with high probability \citep{14}. We project the high dimensional data into a lower dimensional space through a  $p\times k$ random projection matrix ${\mathbb{R}}$, where  $1\le k<n$.
A $p$-dimensional row vector is projected by multiplication on the right by $\mathbb{R}$.
 We make the following assumptions on the random projections.

\medskip\noindent
{\bf Assumption~1.} {\em $ {\mathbb{R}}_{p\times k}$ is a random matrix, independent of the data matrices $\bf{X}$ and $\bf{Y}$, such that ${\mathbb{R}'\mathbb{R}={\bf I}_{k\times k}}$, where $1\le k<n$.}

\medskip\noindent
{\bf Assumption~2.} {\em For any non-zero $p$-dimensional vector $\delta$, the Euclidean norm  $||\delta'\mathbb{R}||_2$ is a continuous random variable with finite second moment.}

\medskip
Assumption~1 implies that the elements of the random projection matrix are not independent. In fact, the matrix is semi-orthogonal. When the elements of the matrix $\mathbb{R}$ are continuous random variables with finite second moment, Assumption~2 is satisfied.

The pooled sample covariance matrix of the projected data matrices $\bf{X\mathbb{R}}$ and $\bf{Y\mathbb{R}}$ is ${\bf{\mathbb{R}}}'S{\bf\mathbb{R}}$.

\medskip\noindent
{\bf Lemma 1.} {\em If Assumption~1 holds and $\Sigma$ is positive definite (denoted by $\Sigma>0$), then ${\bf{\mathbb{R}}}'S{\bf\mathbb{R}}$ is also positive definite (i.e., ${\bf{\mathbb{R}}}'S{\bf\mathbb{R}}>0$) with probability 1.}\\

Hotelling's $T^2$ statistic for the projected data matrices $\bf{X\mathbb{R}}$ and $\bf{Y\mathbb{R}}$ is given as
\begin{eqnarray}\label{randomstat}
T_{\mathbb{R}}^2&= &(n_1\inv + n_2\inv)\inv(\overline{X}-\overline{Y})' {\bf{\mathbb{R}}}({\bf{\mathbb{R}}}'S{\bf\mathbb{R}})^{-1} {\bf{\mathbb{R}}}'(\overline{X}-\overline{Y}).
\end{eqnarray}
In view of Lemma~1, the statistic $T_{\mathbb{R}}^2$ is well defined.

\bigskip\noindent
A randomized extension of Hotelling's $T^2$ test for the hypothesis (\ref{hyp}) is
\begin{eqnarray}\label{randomtest}
\phi(T_\mathbb{R}^2)=\begin{cases} 1&\mbox{if}~ \frac{n-k+1}{k}\, \frac{T^2_\mathbb{R}}{n}>c_\alpha,\\ 0&\mbox{otherwise},\end{cases}
\end{eqnarray}
where $c_\alpha$ is chosen such that
\begin{equation}
P\left[ \frac{n-k+1}{k}  \, \frac{T_{\mathbb{R}}^2}{n}>c_\alpha \bigg| \bf{H_0}\right]=\alpha.\label{dr01}
\end{equation}

\noindent{\bf Theorem 1.} {\em Let $c_\alpha$ be such that $F_{k,n-k+1}(c_\alpha)=1-\alpha$, where $F_{r,s}(\cdot)$ is the $F$-distribution function with numerator and denominator degrees of freedom $r$ and $s$, respectively. If a projection matrix $\mathbb{R}$ satisfies Assumption~1 and $\Sigma>0$, then the following holds.
\begin{enumerate}[(a)]
\item $E\left[\phi(T_\mathbb{R}^2)\big|\bf{H_0}\right]=\alpha.$
\item Let $\bf{H}_1^*$ denote a sequence of alternative hypotheses such  that $n_1$, $n_2$, $p$, and $k$ converge to $\infty$,  $k/n \to c \in (0,1)$, and there is a sequence $\delta \to \infty$ such that
\begin{equation}
E_{\mathbb{R}} P_{X,Y} \left( \sqrt{n} (n_1\inv+n_2\inv)\inv \Delta_\mathbb{R}/k  \ge \delta \bigg| \mathbb{R}, {\bf H}_1^*\right) \to 1,\label{dr04}
\end{equation}
where ${\bf\Delta_\mathbb{R}}=(\mu_1-\mu_2)'\mathbb{R}(\mathbb{R}'\Sigma \mathbb{R})^{-1}\mathbb{R}'(\mu_1-\mu_2)$.
Then, under Assumption~2,
$E[\phi(T_\mathbb{R}^2)|\bf{H^*_1}] \to$~ $1$.
\item Under Assumption~2, $E[\phi(T_\mathbb{R}^2)|\bf{H_1}]\ge\alpha.$
\end{enumerate}}

Let $a\sim b$ mean that $0 < \liminf (a/b) \le \limsup (a/b) < \infty$.
If $n_1 \sim n_2$  and $k/n \to c \in (0,1)$, then $(n_1\inv+n_2\inv)\inv/k \sim 1$, and
then \eqref{dr04} implies that $\sqrt{n} \Delta_\mathbb{R} \to \infty$.  This is a weak assumption as the examples in Section~\ref{sec:assumption6} show.

Theorem~1 (a) and (b) show that the randomized test (\ref{randomtest}) is a consistent exact size $\alpha$ test.  Further, part (c) of Theorem~1 shows that the randomized test (\ref{randomtest}) is an unbiased test. It is important to note that this randomized test does not impose any restriction on the dimension $p$.

\citet{15} showed that the empirical distribution of randomly projected data is close to a Gaussian distribution. Using this fact, the randomized test given above can be adopted even when the data are not Gaussian.

\section{RAPTT}\label{s3}

A single random-projection Hotelling test might have less power than the standard Hotelling test.  Even worse, it could lead to different conclusions in the testing problem (\ref{hyp}) for different realizations of the projection matrix $\mathbb{R}$. To address this issue, we average the p-values of $m$ random-projection Hotelling tests  using independently generated $\mathbb{R}$.

Note that the p-value of random-projection Hotelling test (\ref{randomtest}) is
\begin{eqnarray}\label{pval}
\theta&=&1-F_{ k,n-k+1}\left(\frac{n-k+1}{k}\cdot\frac{T_{\mathbb{R}}^2}{n}\right),
\end{eqnarray}
where $F_{r,s}(\cdot)$ is the $F$-distribution with degrees of freedom $r$ and $s$. (Recall that $n=n_1+n_2-2$.) Let $\mathbb{R}_1^*, \mathbb{R}_2^*,\ldots ,\mathbb{R}_m^*$ be $m$ independent and identically distributed projection matrices.  Let the p-value of the random-projection Hotelling test corresponding to the projection matrix $\mathbb{R}_i^*$ be $\theta_i^*$.

RAPTT is defined as
\begin{eqnarray}\label{randomemptest}
\phi^*=\begin{cases} 1&\mbox{if}~ \bar\theta^*<u_\alpha,\\
 0&\mbox{otherwise},\end{cases}
\end{eqnarray}
where $\bar\theta^*=\frac{1}{m}\sum_{i=1}^m \theta_i^*$
and $u_{\alpha,n_1,n_2}$ is chosen such that $P\left[  \bar\theta^*<u_{\alpha,n_1,n_2} \bigg| {\bf H_0}\right]=\alpha$.\\

\noindent
{\bf Theorem 2.} {\em If the projection matrices $\mathbb{R}_1^*$, $\mathbb{R}_2^*,\ldots$,$\mathbb{R}_m^*,$ satisfy Assumptions~1, 2 and $\Sigma>0$, for  fixed sample sizes $n_1,~n_2$, and projected dimension $k$ and $m\rightarrow\infty$, the distribution of $ \bar\theta^*$ under {\bf $H_0$} does not depend upon the parameter $(\mu_1=\mu_2, \Sigma)$.}\\

In view of Theorem~2, the cutoff $u_\alpha$ in (\ref{randomemptest}) can be computed empirically.   One can simulate the distribution of $\bar\theta^*$ for some arbitrary choice of $\mu_1 = \mu_2$ and $\Sigma$, e.g.,
$\mu_1 =\mu_2 =0$ and $\Sigma = I$.  Conditionally, given the data matrices $\bf{X}$ and $\bf{Y}$, the p-values $\theta_i^*$, for $i=1,2,\ldots,m$, are independent and identically distributed.  Unconditionally, they are of course dependent.  To simulate the null distribution of RAPTT, one simulates $K$ data sets from the null distribution, or, to reduce the computational burden, simulate only the sufficient statistics, $\overline{X}$, $\overline{Y}$, and $S$ . For the $k$th of these data sets (or sets of sufficient statistics), one computes $\bar\theta^*_k$ using $m$ independent random projections.  Then the empirical distribution of $\bar\theta^*_1,\ldots,\bar\theta^*_K$ approximates the null distribution of $\bar \theta^*$ and can be used to compute $u_\alpha$.  RAPTT becomes exact as $K \to \infty$ even for fixed $m$, although we recommend large values for both $K$ and $m$.\\

\noindent{\bf Theorem 3.} {\em If $\Sigma>0$, if the projection matrices $\mathbb{R}_1^*$ , $\mathbb{R}_2^*,\ldots$,$\mathbb{R}_m^*$ satisfy Assumption~1 and 2, and if the assumptions of Theorem 1 (b) hold and $m$ is fixed,  then the  test (\ref{randomemptest}) is consistent, i.e.,
$\lim_{n_1,n_2\rightarrow\infty}E[\phi^*|{\bf H_1^*}]= 1$.}

\section{Choice of $\mathbb{R}$ and $k$}\label{sR}

The building block of RAPTT is the random-projection Hotelling test given by (\ref{randomtest}). Test (\ref{randomtest}) can be applied with any projection matrix $\mathbb{R}$ and any dimension of the projected space $k$ that satisfy Assumptions~1 and~2.  However, the power of the random-projection Hotelling test and of RAPTT will depend on the choice of $\mathbb{R}$ and $k$.

\subsection{Choice of $k$}

If $k\approx n$, one would expect that the power of the test (\ref{randomtest}) would be small in accordance with \citet{BS}. Further, smaller values of $k$  might not adjust properly for correlations in the data; the choice $k=1$ ignores correlation entirely.   We will choose $\mathbb{R}$ and $k$ with the hope that the power of the random-projection Hotelling test (\ref{randomtest}) could be maximized.

From (\ref{powereq}) in the Appendix, the exact power of random-projection Hotelling test is
\begin{eqnarray}
 E[\phi(T_\mathbb{R}^2)|\bf{H_1}]&=&P\left[ \frac{n-k+1}{k}\cdot\frac{T^2_\mathbb{R}}{n}>c_\alpha \bigg|{\bf H_1}\right]\label{powerfirst}\nonumber\\
 &\!\!\!\!=\!\!\!\!\!\!&1\!-\!E_\mathbb{R}\left\{\sum_{l=0}^{\infty}\frac{e^{-\frac{n_1n_2}{n_1\!+\!n_2}\frac{{\bf\Delta_\mathbb{R}}}2}\left(\frac{n_1n_2}{n_1\!+\!n_2}\frac{{\bf\Delta_\mathbb{R}}}2\right)^l}{l!}I_{\frac{kc_\alpha}{kc_\alpha\!+n-\!k\!+\!1}}\left(\frac{k\!+\!2l}{2},\frac{n-\!k\!+\!1}{2}\right)\right\},\label{powersecond}
\end{eqnarray}
where, as before, ${\bf\Delta_\mathbb{R}}=(\mu_1-\mu_2)'\mathbb{R}(\mathbb{R}'\Sigma \mathbb{R})^{-1}\mathbb{R}'(\mu_1-\mu_2)$, and the function $I$ is the regularized incomplete beta function given by (\ref{incompletebeta}) in the Appendix.
Note that the power (\ref{powersecond}) depends on $k$ and $\mathbb{R}$ explicitly through ${\bf \Delta_\mathbb{R}}$ and $I$. It is important to emphasize that the power expression given by (\ref{powersecond}) also depends upon the unknown parameter $\Sigma$, so  maximizing the power by selecting the optimal $\mathbb{R}$ and $k$ appears to be a rather challenging task.

It can be seen from (\ref{powerfirst}) and (\ref{incompletebeta}) that for fixed ${\bf\Delta_\mathbb{R}}$ and $k$, the power would be the largest when $c_\alpha$ is smallest. Recall $c_\alpha$ is the upper quantile of $F$ distribution with degrees of freedom $k$ and $n-k+1$. We choose  the $k$ that minimizes $c_\alpha$ over $k$.
In Section~\ref{s4}, we observe that the empirical power of the test (\ref{randomtest}) corresponding to this intuitive choice of $k$ is very close to the empirical optimal power of the test under the simulation set-up.

\subsection{Choice of $\mathbb{R}$}

We now turn to the choice of projection matrix $\mathbb{R}$. A natural choice $\mathbb{R}$ is to draw random matrices uniformly on the set of $p\times k$ dimensional real matrices such that $\mathbb{R}'\mathbb{R}=I$, i.e., choose the projection matrix from the Haar distribution on this set of real matrices. A projection matrix generated in this manner satisfies Assumptions~1 and~2. We denote this choice by $\mathbb{R}^1$.

Our second choice of $\mathbb{R}$ is based on the idea of {\em one permutation + one random projection}, which is closely related to  {\em very sparse random projection}~\citep{Li06} and {\em count-sketch}~\citep{Article:Charikar_2004}. Let $[r_1, r_2, \ldots, r_p]$ be a vector of i.i.d.\ absolutely continuous random variables with finite second moment. Without loss of generality, we assume the dimensionality $p$ is divisible by $k$, and we break the $n \times p$ data matrix's columns (i.e., variables)  evenly  into $k$ blocks. We conduct one random projection on the first block (i.e., data matrix columns 1 to $p/k$) using weights (i.e., projection vector) $[r_1, r_2, \ldots, r_{p/k}]$, then on the second block (i.e., columns $p/k+1$ to $2p/k$) using a projection vector $[r_{p/k+1}, \ldots r_{2p/k}]$, and so on. This way, we still obtain a projected data matrix of $k$ columns. To remove  the influence of the structure of data, we first randomly  permute the columns of the original (non-projected) data matrix before we break the columns into $k$ blocks.

Equivalently, we can view the second choice as a random projection  matrix $\mathbb{R}$ of size $p\times k$. Here, we provide the following simple example of $\mathbb{R}$ for $p=4$ and $k=2$:

\begin{align}\notag
\left[\begin{array}{c}
r_1\\
r_2\\
r_3\\
r_4
\end{array}
\right]\overset{blocking}{\Longrightarrow }
\left[\begin{array}{cc}
r_1&0\\
r_2&0\\
0&r_3\\
0&r_4
\end{array}
\right]\overset{permutation}{\Longrightarrow}
\left[\begin{array}{cc}
0&r_4\\
r_1&0\\
r_2&0\\
0&r_3
\end{array}
\right]\overset{normalization}{\Longrightarrow}\mathbb{R}'=
\left[\begin{array}{cc}
0&\frac{r_4}{\sqrt{r_3^2+r_4^2}}\\
\frac{r_1}{\sqrt{r_1^2+r_2^2}}&0\\
\frac{r_2}{\sqrt{r_1^2+r_2^2}}&0\\
0&\frac{r_3}{\sqrt{r_3^2+r_4^2}}
\end{array}
\right]
\end{align}
See the analysis  by \citet{Li11} in the context of using this type of projection matrix for estimating massive data pairwise inner products, where  $r_i$ is restricted to the sub-Gaussian family.

\subsection{On Condition \eqref{dr04}}\label{sec:assumption6}

Condition \eqref{dr04} is used in the proof of Theorem 1 to ensure that the difference between the mean and the critical value of the test statistic is a larger order of magnitude compared to the test statistic's standard deviation.

To explore this assumption, we will consider the simple case where $n_1=n_2$ and $k = cn$ and $p = Mn$, where $0 < c <1 < M$ and $M/c = p/k$ is an integer.  For simplicity, we will also assume that $\Sigma = {\bf I}_p$, the $p \times p$ identity matrix, and that $\mathbb{R}$ is of the second type, that is, one permutation and one random projection.
Thus, before the permutation
\[
\mathbb{R} = \left(\begin{matrix} b_1 & 0 & \cdots & 0 \\
0 & b_2 & \cdots &0 \\
\vdots & \vdots & \ddots & \vdots \\
0 & 0 & \cdots & b_{k}
\end{matrix}
\right)
\]
where each of $b_1,\cdots,b_{k}$ is a column vector containing $M/c$ iid $r_i$.
For the present analysis, the permutation is irrelevant and will be ignored.
It follows that $\mathbb{R}'\Sigma\mathbb{R} \approx m_2 \, M/c  \,  {\bf I}_{k} $
where $m_2$ is the second moment of $r_i$.

First, suppose that $\mu_1 - \mu_2= d {\bf 1}_p$ where $d$ is a scalar depending on $n$ and ${\bf 1}_p$ is a $p$-dimensional vector of ones.  Then $\mathbb{R}'(\mu_1-\mu_2)
\approx  d m_1 M/c {\bf 1}_{k}$ where $m_1$ is the mean of $r_i$, which we will assume is not zero.  Then
\begin{equation}
\Delta_\mathbb{R} \approx \frac{(d m_1)^2 {\bf 1}_{k}' {\bf I}_{k} {\bf 1}_{k}} {m_2} \sim k \, d^2 \sim n \, d^2.
\label{dr05}
\end{equation}
With these choices of $n_1$, $n_2$, and $k$, \eqref{dr04} will hold if $\sqrt n \Delta_\mathbb{R} \to \infty$.
It  then follows from \eqref{dr05} that \eqref{dr04} holds if $n^{3/4} d \to \infty$, so $d$ could converge to 0 quite slowly and still have consistency.  In summary, detecting that $\mu_1$ and $\mu_2$ differ by a fixed amount at every coordinate is relatively easy and that fixed difference can be small.
If we now assume that $M<1$ but keep the other assumptions unchanged including that $p < M n$,
then the the Hotelling T-test is defined.  Calculations similar to those just completed show that the Hotelling test is also consistent if $n^{3/4} d \to \infty$.  This result suggests that the random projection Hotelling test is competitive with the Hotelling test itself.

Next, suppose that $\mu_1 - \mu_2=d {\bf e}_1$ where ${\bf e}_1$ is the unit vector (a one followed by $p-1$ zeros), but, otherwise, let $n_1$, $n_2$, $p$, $k$, and $\Sigma$ be as before.  One can show that $\Delta_\mathbb{R} \sim d^2/n$
and then \eqref{dr04} holds if $ n^{-1/4} d \to \infty$, so that $d$ must converge to $\infty$ at a rate greater than $n^{1/4}$ for consistency.  (Thus, detecting that $\mu_1$ and $\mu_2$ differ only at a single coordinate is like searching for a needle in a haystack---we need a big needle.)

For comparison, suppose $p$ is fixed and a Hotelling's T-test is used.  Suppose also that $\mu_1-\mu_2 = d \, \bf{e}$ for $d$ depending on $n$ and $\bf{e}$ a fixed non-zero vector.  That $n^{1/2} d \to \infty$ is sufficient for consistency.

\section{Simulation of Performance}\label{s4}
In this section,
we consider the finite sample performance of RAPTT and  compare it to that of the asymptotic tests mentioned in Section~\ref{s1}.   First, we briefly describe three major competing tests.

\subsection{Competing Tests}

\def\tr{{\textrm{tr}}}

\citet{BS} considered the statistic
\begin{eqnarray}
BS&=&\frac{\frac{n_1n_2}{n_1+n_2} (\overline{X}-\overline{Y})' (\overline{X}-\overline{Y})-\tr(S)}{\sqrt{\frac{2n(n+1)}{(n+2)(n-1)}\left[\tr(S^2)-\frac1n(\tr{S})^2\right]}},
\end{eqnarray}
where $\tr(A)$ is the trace of the matrix $A$. The BS test rejects hypothesis (\ref{hyp}) if $BS\ge z_\alpha$ where $z_\alpha$ is the $1-\alpha$ quantile of the standard normal distribution.

The modified statistic proposed by \citet{CQ} is
\begin{eqnarray}
CQ&=&\frac{\frac{\sum_{i\ne j}X_iX_j'}{n_1(n_1-1)}+\frac{\sum_{i\ne j}Y_iY_j'}{n_1(n_1-1)}-2\frac{\sum_{i=1}^{n_1}\sum_{j=1}^{n_2}X_iY_j'}{n_1n_2}}{\hat{\sigma}_n},
\end{eqnarray}
where $\hat{\sigma}_n$ is an estimate of standard error of the numerator. (For the formula, see \citet{CQ}.)  The CQ test rejects the hypothesis (\ref{hyp}) if $CQ\ge z_\alpha$.

\citet{SD} considered the statistic
\begin{eqnarray}
SD&=&\frac{\frac{n_1n_2}{n_1+n_2} (\overline{X}-\overline{Y})'[\Diag(S)]^{-1} (\overline{X}-\overline{Y})-\frac{np}{n-2}}{\sqrt{2\left(\tr(R^2)-\frac{p^2}{n}\right)\left(1+\frac{\tr(R^2)}{p^{3/2}}\right)}},
\end{eqnarray}
where  $R=\Diag(S)^{-\frac12}S \,  \Diag(S)^{-\frac12}$. The SD test rejects hypothesis (\ref{hyp}) if $SD\ge z_\alpha$.

The asymptotic superiority of one of these tests over the others depends upon the structure of the covariance matrix $\Sigma$. For example, if $\Sigma$ is a diagonal matrix, then the SD test has larger asymptotic power than that of the other tests.
If $p\gg n$, then the CQ test has larger asymptotic power than the others (see \citet{SD,CQ}).

\subsection{Covariance Matrices}

We consider the following four covariance matrices for the simulation study.
\begin{itemize}
\item $\Sigma_1={\bf I}$.
\item $\Sigma_2=\Diag(\lambda_1,\ldots,\lambda_p)$ where $\lambda_i=\frac{20}{i}$ for $i=1,\ldots,20$ and $\lambda_i=1$ for $i=21,\ldots,p$.
\item $\Sigma_3$ is a symmetric Toeplitz matrix generated with  $(\eta_1,\eta_2,\ldots,\eta_p)$ where $\eta_1=1$, $\eta_2=0.4$,  $\eta_i=0$ for $i=3,\ldots,p$. This corresponds to the covariance matrix of an MA(2) time series.
\item $\Sigma_4$ is a block diagonal matrix with blocks $B$ of size 25, where $B=0.85\times{\bf I}+0.15\times{\bf 1}{\bf 1'}$.
\end{itemize}

\subsection{Alternatives}

We consider a natural alternative for the mean difference together with the alternative chosen by \citet{CQ}. Without loss of generality, we let $\mu_1=0$. Further, we let $1\%$, $5\%$, $25\%$, $50\%$ and $75\%$ of the $p$ coordinates of $\mu_2$  be non zero. The non-zero coordinates of $\mu_2$ are chosen randomly with equal probability for each level of mean difference.
\begin{itemize}
\item {\it Alternative 1}: Non-zero elements of $\mu_2$ are $N(1,1)$ rescaled such that\\ $\frac12(\mu_1-\mu_2)'\Sigma^{-1}(\mu_1-\mu_2)=1.$
\item {\it Alternative 2}: Non-zero elements of $\mu_2$ are $N(1,1)$ rescaled such that $\frac{||\mu_1-\mu_2||^2}{\sqrt{\tr(\Sigma^2)}}=0.1$. This is the
alternative hypothesis used in the simulation study of \citet{CQ}
\end{itemize}

\subsection{Random Projection Matrices and Empirical Null Distributions}

 We choose two different random projection matrices: $\mathbb{R}^1$ as Haar distributed and $\mathbb{R}^2$ obtained from {\em one permutation + one random projection} as described in Section~\ref{s3}. We choose the dimensions $p=200$ and $p=1000$ to illustrate the performance in high dimensions. We choose $n_1=n_2=50$ for the dimension $p=200$. For $p=1000$, we choose $n_1=n_2=70$. The projected dimension $k$ is chosen as described in Section~\ref{s3}, and is $k=43$ for $p=200$ and $k=62$ for $p=1,000$.

Figure~1 shows the plot of the empirical null distribution of $\bar\theta^*$ based on 1,000 simulated samples from $H_0$ and, for each such data set, 5,000 random projections for all the choices of dimension, projection matrices, as well as covariance matrices. The plots  indicate that the null distribution does not depend upon the choice of the covariance matrix, in agreement with Theorem~2. From the columns of Figure~1, it appears that the empirical distribution corresponding to the projection matrices $\mathbb{R}^1$ and $\mathbb{R}^2$ are similar to each other indicating invariance, or at least near invariance, over the choice of the projection matrices.

\begin{figure}[h!]
\begin{center}
\mbox{
\includegraphics[width=2.6in]{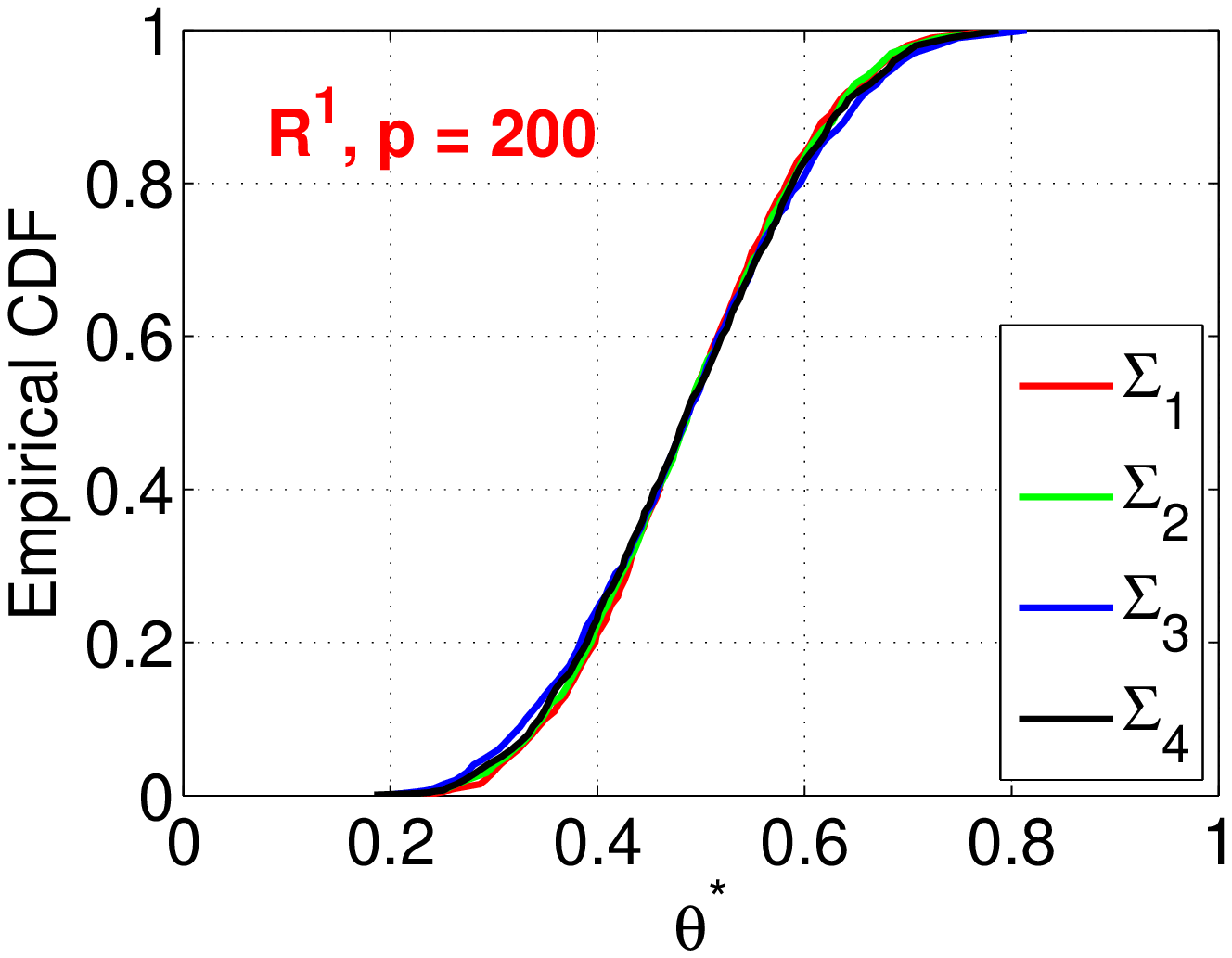}\hspace{0.2in}
\includegraphics[width=2.6in]{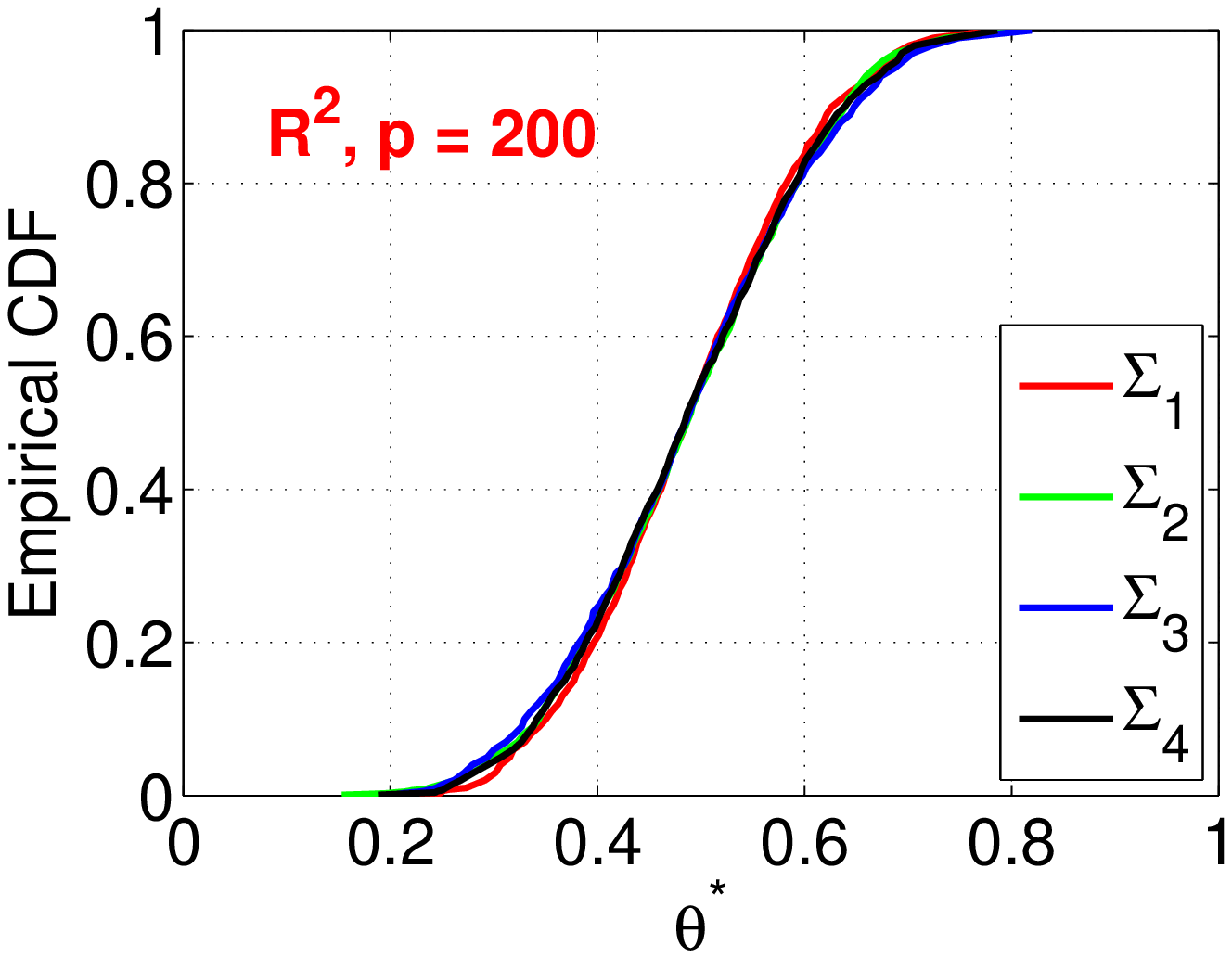}
}
\mbox{
\includegraphics[width=2.6in]{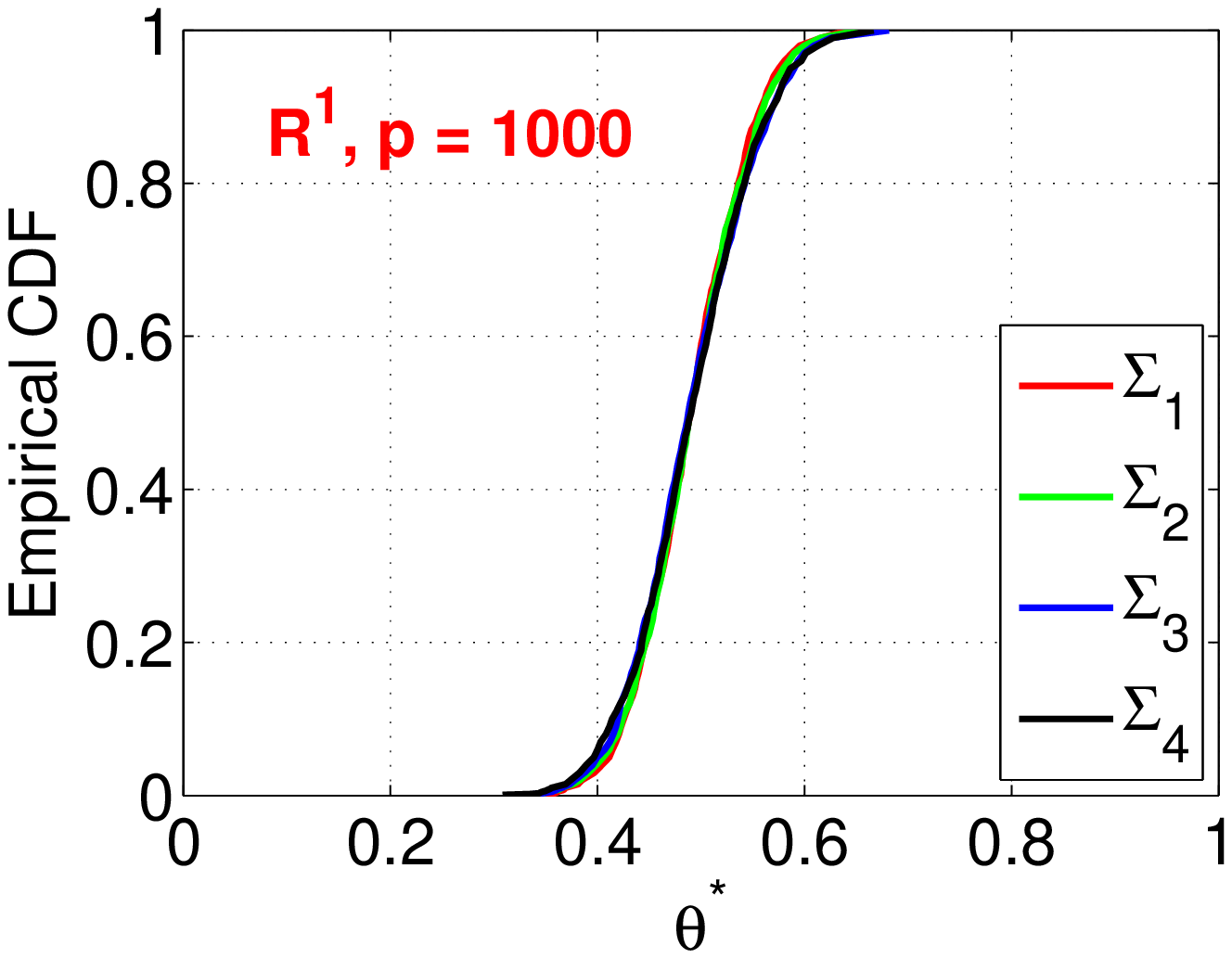}\hspace{0.2in}
\includegraphics[width=2.6in]{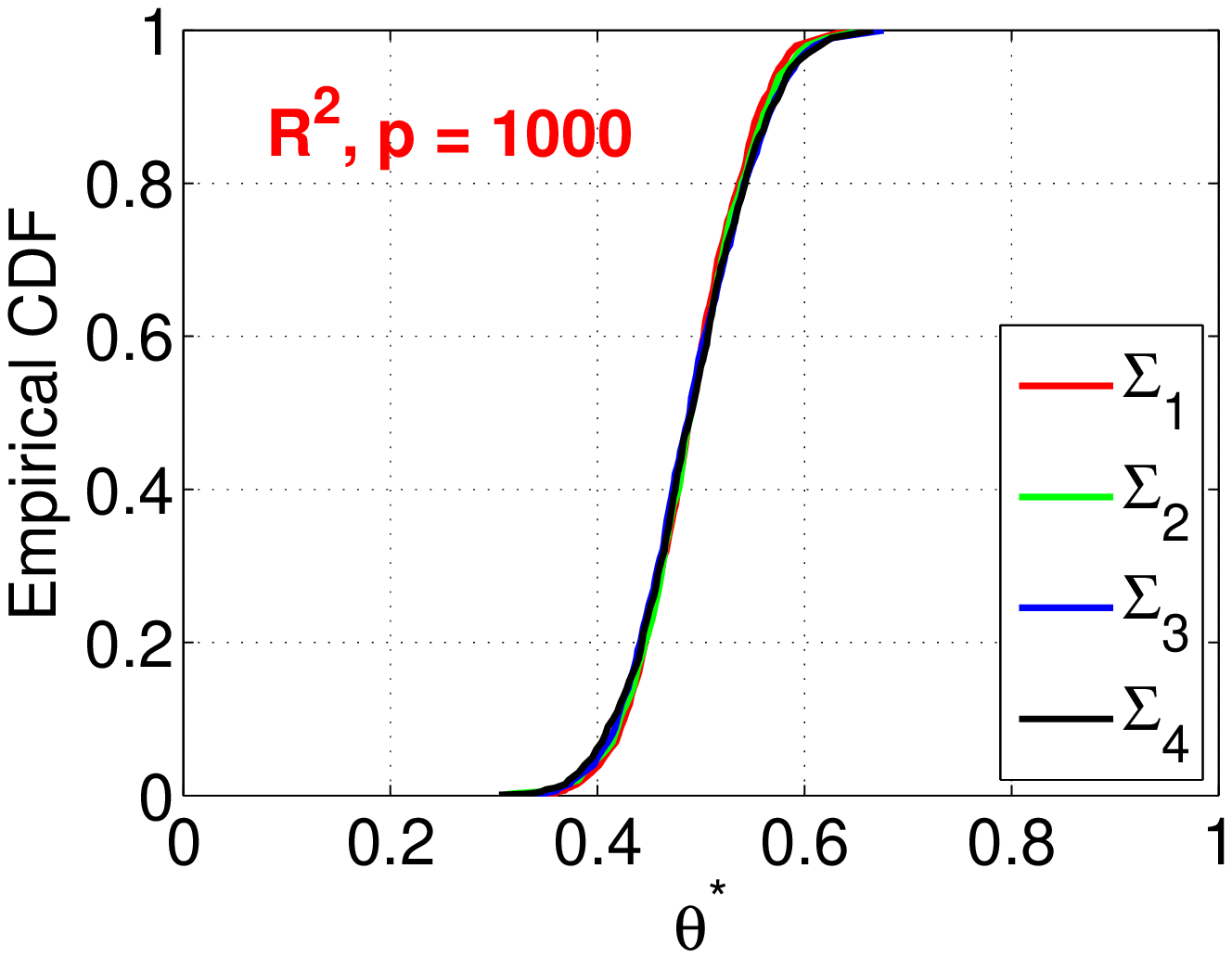}
}
\vspace{-0.1in}
\caption{\small Plots of empirical null distribution of $\bar\theta^*$ for $n_1=n_2=50; p=200$ (first row) and $n_1=n_2=70; p=1000$ (second row) based on 1000 simulation runs and $m=5000$ bootstrap samples, first and second column corresponds to projection matrices $\mathbb{R}^1$ and $\mathbb{R}^2$ respectively.}
\end{center}\vspace{-0.2in}
\end{figure}

The empirical cutoff for the proposed bootstrap test is computed on the basis of the combined empirical null distribution corresponding to the different covariance matrices. 
The empirical power is computed from 1,000 simulation runs.

\subsection{Empirical Significance Levels and Powers}

Tables 1, 2, 3 and 4 show the empirical significance level and power of the test for the four choices of the true covariance matrix, the nominal level of significance being 0.05. The last rows of the four tables indicate that the empirical significance level of the proposed test in each case is very close to the true significance level, relative to Monte Carlo error.  Assuming that the true significance level is 0.05, the approximate Monte Carlo standard error is $\sqrt{(0.05)(0.95)/1000} =$ 0.0069.

We first compare the empirical power of the proposed test with that of the other tests, starting with the BS and the CQ tests. For $\Sigma_1$, i.e., the covariance matrix being identity matrix, Table~1 indicates that the empirical power of RAPTT is smaller than that of the two existing tests for $p=200$ and is marginally smaller than them for $p=$ 1,000, for all choices of alternatives. Tables~2 shows that, for the covariance matrix $\Sigma_2$ the empirical power of RAPTT is much larger than those of the CQ and BS tests for both choices of dimension and both  alternatives. For $\Sigma_3$ and $\Sigma_4$, Table~3 and~4 show that the empirical power of RAPTT is comparable to those of the CQ and BS tests for dimension $p=200$. Further, for $p=1000$ RAPTT has larger power. In summary, RAPTT has larger power than those of the CQ and BS tests for  the choices of $\Sigma$ and alternatives, when p=1000. For $p=200$, the power of RAPTT is either larger than or comparable to those of the CQ and  BS tests.

\begin{table}[h!]
\def~{\hphantom{0}}
\caption{Empirical power and size corresponding to $\Sigma_1$. $\mathbb{R}^1$ and $\mathbb{R}^2$ are RAPTT with the two choices of random projection matrix.\vspace{-0.2in}}
{\footnotesize
\begin{tabular}{ll|ccccc|ccccc}
 \\
\hline
 \multicolumn{2}{r}{ Non-zero $\%$ of }&\multicolumn{5}{c}{p=200, $n_1=n_2=50$} &\multicolumn{5}{c}{p=1000, $n_1=n_2=70$} \\
 &$\mu_1-\mu_2$ &$\mathbb{R}^1$&$\mathbb{R}^2$&CQ&SD&BS &$\mathbb{R}^1$&$\mathbb{R}^2$&CQ&SD&BS \\
\hline
Alt. 1 &$1\%$ &0.614 &0.485 &0.739 &0.687 &0.740   &0.409 &0.420 &0.467 &0.414 &0.467 \\
&$5\%$ & 0.619 &0.601 &0.744 &0.707 &0.745   &0.399 &0.420 &0.437 &0.391 &0.460  \\
&$25\%$& 0.619&0.627 &0.759 &0.717 &0.738   &0.387 &0.407 &0.449 &0.389 &0.449 \\
&$50\%$ & 0.630 &0.625 &0.753 &0.717 &0.753   &0.403 &0.422 &0.467 &0.405 &0.468 \\
&$75\%$ & 0.630&0.636 &0.760 &0.728 &0.760   & 0.403& 0.418& 0.466& 0.400& 0.466\\
\hline
Alt. 2 &$1\%$ &0.409 &0.397 &0.518 &0.481 &0.518   &0.728 &0.731 &0.778 &0.727 &0.778 \\
&$5\%$ &0.389 &0.396 &0.511 &0.470 &0.512   &0.732 &0.729 &0.783 &0.737 &0.778  \\
&$25\%$& 0.387 & 0.320&0.514 &0.483 &0.514   &0.722 &0.750 &0.789 &0.742 &0.789  \\
&$50\%$ & 0.392&0.393 &0.514 &0.482 &0.514   &0.709 &0.725 &0.766 &0.719 &0.764  \\
&$75\%$ & 0.407& 0.426&0.524 &0.491 &0.522   &0.718 &0.735 &0.774 &0.721 &0.771  \\
\hline\
Null&$0\%$&0.034 &0.040 &0.062 &0.058 &0.062  &0.039  &0.037 &0.051 &0.041 &0.051 \\
\hline
\end{tabular}
\label{table1}}\vspace{0.15in}
\end{table}

\begin{table}[h!]
\def~{\hphantom{0}}
\caption{Empirical power and size corresponding to $\Sigma_2$.   $\mathbb{R}^1$ and $\mathbb{R}^2$ are RAPTT with the two choices of random projection matrix.\vspace{-0.2in}}
{\footnotesize
\begin{tabular}{ll|ccccc|ccccc}
 \\
\hline
 \multicolumn{2}{r}{ Non-zero $\%$ of }&\multicolumn{5}{c}{p=200, $n_1=n_2=50$} &\multicolumn{5}{c}{p=1000, $n_1=n_2=70$} \\
 &$\mu_1-\mu_2$ &$\mathbb{R}^1$&$\mathbb{R}^2$&CQ&SD&BS &$\mathbb{R}^1$&$\mathbb{R}^2$&CQ&SD&BS \\
\hline
Alt. 1 &$1\%$ &0.539 &0.455 &0.300 &0.687 &0.300   &0.423 &0.434 &0.343 &0.413 &0.343 \\
&$5\%$ &0.551 &0.561 &0.323 &0.718 &0.323   &0.389 &0.405 &0.319 &0.391 &0.319  \\
&$25\%$&0.548 &0.584 &0.311 &0.725 &0.312   &0.396 &0.406 &0.317 &0.387 &0.317 \\
&$50\%$&0.587 &0.617 &0.341 &0.721 &0.339   &0.392 &0.420 &0.330 &0.404 &0.330 \\
&$75\%$&0.586 &0.619 &0.334 &0.724 &0.334   &0.385 &0.413 &0.325 &0.404 &0.327 \\
\hline
Alt. 2 &$1\%$&0.787 &0.633 &0.499 &0.905 &0.497   &0.853 &0.839 &0.764 &0.855 &0.764 \\
&$5\%$ &0.767 &0.768 &0.473 &0.879 &0.475   &0.847 &0.870 &0.776 &0.885 &0.776  \\
&$25\%$&0.803 &0.828 &0.498 &0.912 &0.499   &0.845 &0.870 &0.784 &0.888 &0.786  \\
&$50\%$&0.782 &0.812 &0.489 &0.890 &0.489   &0.839 &0.864 &0.766 &0.869 &0.765  \\
&$75\%$&0.791 &0.809 &0.501 &0.894 &0.499   &0.833 &0.860 &0.770 &0.868 &0.770  \\
\hline
Null&$0\%$& 0.055&0.047 &0.075 &0.058 &0.075   &0.047 &0.044 &0.047 &0.041 &0.048 \\
\hline
\end{tabular}
\label{table2}}
\vspace{0.15in}
\end{table}

\begin{table}[h!]
\def~{\hphantom{0}}
\caption{Empirical power and size corresponding to $\Sigma_3$.  $\mathbb{R}^1$ and $\mathbb{R}^2$ are RAPTT with the two choices of random projection matrix.\vspace{-0.2in}}
{\footnotesize
\begin{tabular}{ll|ccccc|ccccc}
 \\
\hline
 \multicolumn{2}{r}{ Non-zero $\%$ of }&\multicolumn{5}{c}{p=200, $n_1=n_2=50$} &\multicolumn{5}{c}{p=1000, $n_1=n_2=70$} \\
 &$\mu_1-\mu_2$ &$\mathbb{R}^1$&$\mathbb{R}^2$&CQ&SD&BS &$\mathbb{R}^1$&$\mathbb{R}^2$&CQ&SD&BS \\
\hline
Alt. 1 &$1\%$&0.377 &0.311 &0.364 &0.341 &0.364   &0.248 &0.258 &0.218 &0.179 &0.218 \\
&$5\%$&0.363 &0.357 &0.381 &0.339 &0.382   &0.237 &0.251 &0.205 &0.161 &0.204  \\
&$25\%$&0.469 &0.475 &0.487 &0.437 &0.486   &0.241 &0.251 &0.219 &0.174 &0.220 \\
&$50\%$&0.436 &0.441 &0.434 &0.401 &0.434   &0.269 &0.278 &0.242 &0.205 &0.242 \\
&$75\%$&0.486 &0.493 &0.483 &0.545 &0.485   &0.311 &0.334 &0.292 &0.239 &0.292 \\
\hline
Alt. 2 &$1\%$& 0.525 &0.425 &0.518 &0.481 &0.518   &0.798 &0.797 &0.771 &0.730 &0.771 \\
&$5\%$ &0.519 &0.503 &0.518 &0.472 &0.518   &0.800 &0.805 &0.776 &0.715 &0.776  \\
&$25\%$&0.495 &0.497 &0.517 &0.472 &0.517   &0.816 &0.823 &0.783 &0.735 &0.781  \\
&$50\%$&0.515 &0.517 &0.515 &0.474 &0.515   &0.789 &0.798 &0.765 &0.719 &0.765  \\
&$75\%$&0.509 &0.508 &0.511 &0.473 &0.510   &0.789 &0.796 &0.764 &0.716 &0.766  \\
\hline
Null&$0\%$& 0.063 &0.065 &0.069 &0.052 &0.069   &0.051 &0.052 &0.041 &0.027 &0.041 \\
\hline
\end{tabular}
\label{table3}}
\end{table}

\vspace{0.2in}

\begin{table}[h!]
\def~{\hphantom{0}}
\caption{Empirical power and size corresponding to $\Sigma_4$.  $\mathbb{R}^1$ and $\mathbb{R}^2$ are RAPTT with the two choices of random projection matrix.\vspace{-0.2in}}
{\footnotesize
\begin{tabular}{ll|ccccc|ccccc}
 \\
\hline
 \multicolumn{2}{r}{ Non-zero $\%$ of }&\multicolumn{5}{c}{p=200, $n_1=n_2=50$} &\multicolumn{5}{c}{p=1000, $n_1=n_2=70$} \\
 &$\mu_1-\mu_2$ &$\mathbb{R}^1$&$\mathbb{R}^2$&CQ&SD&BS &$\mathbb{R}^1$&$\mathbb{R}^2$&CQ&SD&BS \\
\hline
Alt. 1 &$1\%$ &0.580 &0.466 &0.540 &0.480 &0.540   &0.336 &0.368 &0.275 &0.234 &0.274 \\
&$5\%$ &0.581 &0.550 &0.534 &0.490 &0.534   &0.385 &0.395 &0.309 &0.248 &0.309  \\
&$25\%$&0.594 &0.592 &0.589 &0.536 &0.589   &0.406 &0.420 &0.329 &0.283 &0.321 \\
&$50\%$&0.629 &0.643 &0.636 &0.580 &0.636   &0.447 &0.465 &0.381 &0.330 &0.381 \\
&$75\%$&0.683 &0.690 &0.715 &0.670 &0.715   &0.529 &0.548 &0.463 &0.401 &0.463 \\
\hline
Alt. 2 &$1\%$ &0.580 &0.465 &0.540 &0.480 &0.540   &0.850 &0.853 &0.786 &0.714 &0.786 \\
&$5\%$ &0.569 &0.543 &0.524 &0.479 &0.526   &0.855 &0.854 &0.779 &0.719 &0.779  \\
&$25\%$&0.507 &0.520 &0.500 &0.472 &0.500   &0.842 &0.853 &0.775 &0.720 &0.775  \\
&$50\%$&0.505 &0.507 &0.498 &0.462 &0.498   &0.815 &0.818 &0.753 &0.689 &0.753  \\
&$75\%$&0.472 &0.479 &0.509 &0.466 &0.509   &0.776 &0.789 &0.742 &0.693 &0.741  \\
\hline
Null&$0\%$& 0.050&0.046 &0.079 &0.064 &0.079   &0.059 &0.057 &0.050 &0.033 &0.050 \\
\hline
\end{tabular}
\label{table4}}
\end{table}


We now compare the empirical power of RAPTT with the SD test. For $\Sigma_1$ and $p=200$, Table~1 shows that the power of RAPTT is slightly less than that of the SD test, while for $p=1000$, the power of the two tests is comparable. For $\Sigma_2$ and $p=200$, Table~2 shows that the power of SD test is larger than that of \mbox{RAPTT}. However, for $p=1000$, the power of RAPTT is comparable to  that of  SD test. This indicates that RAPTT is comparable or only slightly worse compared to the SD test when the true dispersion matrix is indeed diagonal (i.e., most favorable to the SD test).
For $\Sigma_3$ and $\Sigma_4$, Tables~3 and~4 show that the power of RAPTT is larger for the choices of dimension and  alternatives.

\subsection{Performance of Chosen $k$}

We now turn to the assessment of the appropriateness of the choice of the projected dimension $k$ proposed in Section~\ref{sR}. We use the  same four covariance matrices and Alternative 1. By searching over different values of $k$, one can determine the largest possible power of the proposed test. Using this power as the benchmark, one can compute the relative power of the proposed test when $k$ is chosen as described in Section~\ref{sR}.

Tables~5 and~6 show the ratio between the empirical power of test (\ref{randomtest}), based on 5,000 runs, corresponding to the choice made in Section~\ref{sR} and the empirical optimal power of (\ref{randomtest}) for two choices of the projection matrices: $\mathbb{R}^1$ and $\mathbb{R}^2$.  These tables show that the ratio is greater than $0.85$ for almost all the choices of $\Sigma$ and the dimension $p$, and greater than 0.9 for  majority of the choices. This set of experiments helps verify the proposed method of choosing $k$.

\begin{table}[h!]
\def~{\hphantom{0}}
\caption{Ratio between the power corresponding to recommended $k$ and optimal power using a significance level $\alpha=0.05$ and $\mathbb{R}^1$.}
{\footnotesize
\begin{tabular}{l|ccccc|ccccc}
\hline
 {Covariance}& \multicolumn{5}{c}{ p=200, $n_1=n_2=50$ }&\multicolumn{5}{c}{p=1000 $n_1=n_2=70$} \\
{matrix}& \multicolumn{5}{c}{ Non-zero $\%$ of $\mu_1-\mu_2$ }&\multicolumn{5}{c}{ Non-zero $\%$ of $\mu_1-\mu_2$} \\
 &$1\%$ & $5\%$ &$25\%$ &$50\%$ &$75\%$ &$1\%$ &$5\%$ &$25\%$ &$50\%$ &$75\%$\\
\hline
$\Sigma_1$ &$0.9236$ &$0.9796$ &$0.9213$ &$0.9658$ &$0.8696$ &$0.8744$ &$0.8845$ &$0.8763$ &$0.9427$ &$0.9474$\\
$\Sigma_2$  &0.9240 &0.9286 &0.9142&0.9383&0.9800&0.8515&0.8161&0.8668&0.8967&0.8592\\
 $\Sigma_3$&$0.9375$ &$0.8834$ &$0.9892$ &$0.9949$ &$0.9535$ &$0.8692$ &$0.9026$ &$0.8427$ &$0.8950$ &$0.9526$\\
 $\Sigma_4$ & 0.9325& 0.8873&0.8912&0.9363&0.9737&0.8970&0.8952&0.9401&0.8628&0.8915\\
\hline
\end{tabular}}\vspace{0.2in}
\end{table}

\begin{table}[h!]
\def~{\hphantom{0}}
\caption{Ratio between the power corresponding to recommended $k$ and optimal power, using a significance level $\alpha=0.05$ and $\mathbb{R}^2$.}
{\footnotesize
\begin{tabular}{l|ccccc|ccccc}
\hline
Covariance& \multicolumn{5}{c}{ p=200, $n_1=n_2=50$ }&\multicolumn{5}{c}{p=1000 $n_1=n_2=70$} \\
matrix& \multicolumn{5}{c}{ Non-zero $\%$ of $\mu_1-\mu_2$ }&\multicolumn{5}{c}{ Non-zero $\%$ of $\mu_1-\mu_2$} \\
 &$1\%$ & $5\%$ &$25\%$ &$50\%$ &$75\%$ &$1\%$ &$5\%$ &$25\%$ &$50\%$ &$75\%$\\
\hline
$\Sigma_1$ &$0.9674$ &$0.9470$ &$0.9117$ &$0.8925$ &$0.9967$ &$0.8929$ &$0.8531$ &$0.8345$ &$0.9182$ &$0.9079$\\
$\Sigma_2$ &0.9657 &0.9855 &0.9907&0.9861&1&0.9418 &0.8900&0.8877&0.8773&0.8921\\
 $\Sigma_3$&$0.9520$ &$0.9597$ &$0.9558$ &$0.9564$ &$0.9502$ &$0.9056$ &$0.9096$ &$0.8981$ &$0.8785$ &$0.8943$\\
 $\Sigma_4$ & 0.9308&0.9210 &0.8460&0.9800&0.9954&0.9281&0.9580&0.9175&0.9205&0.9171\\
\hline
\end{tabular}}
\end{table}

\section{Data Analysis}\label{s6}

We consider gene expression data corresponding to $n_1=40$ cases of tumor colon tissue and $n_2=22$ cases of normal colon tissue probed by oligonucleotide arrays\footnote{ \url{http://genomics-pubs.princeton.edu/oncology/affydata/index.html}}.
The data contains the expression of $p=2000$ genes with highest minimal intensity across the $n_1+n_2=62$ tissues. The gene intensity is derived from the 20 feature pairs that correspond to the gene on the chip, derived using the filtering process; see \citep{Alon} for more details. We will use the log transformed data. We apply the proposed bootstrap test based on the projection matrix $\mathbb{R}^{1}$ as well as $\mathbb{R}^{2}$. The empirical cutoff for the bootstrap test (\ref{randomemptest}) corresponding to $5\%$ level of significance turns out to be $0.4259$ based on 10,000 data sets simulated from the null distribution and $m=5,000$ bootstrap samples. The value of test statistic $ \bar\theta^*$ corresponding to $\mathbb{R}^{1}$ and $\mathbb{R}^2$ turns out to be $0.0045$ and $0.0046$. The hypothesis is rejected and the p-values turn out to be 0.

The BS test statistic is $2.8189$ and the corresponding p-value is $0.0024$. Thus, the BS test also rejects the hypothesis. However, the CQ and SD test statistics are $1.3299$ and $0.6696$ with corresponding p-values of $0.0918$ and $0.2516$ leading to non-rejection.

Testing the hypothesis would have been more challenging if the sample size had been even smaller.  As an illustration, we randomly chose $50\%$ of each sample and recomputed the p-values. We repeated this exercise independently 100 times. The median p-values for RAPTT  using $\mathbb{R}^1$ and $\mathbb{R}^2$ and for the BS, CQ and SD tests were  0, 0, 0.1050, 0.3279 and 0.3900, respectively. The exercise was repeated with random subsamples of only $25\%$. The median  p-values for the proposed test with $\mathbb{R}^1$ and $\mathbb{R}^2$, and for the BS, CQ and SD tests were 0, 0, 0.2949, 0.4474 and 0.4653, respectively. Thus, at least in this example, RAPTT rejects the null hypothesis at sample sizes that are too small for competing tests to reject.

\section{Conclusion}\label{s5}
In this paper, we proposed an exact test, called RAPTT, of the equality of the means of two normal populations based on a random projection of Hotelling's $T^2$ test.   The critical value for RAPTT requires that we simulate data under the null distribution.   The empirical study in Section~\ref{s4} indicates that the power of the proposed test can be often larger than that of competing tests, depending upon the structure of $\Sigma$.
The gene expression data analysis in Section~\ref{s6} illustrates that, in practice, RAPTT can work well  compared to  competing asymptotic tests in ``large $p$, small $n$'' situations.

\vspace{0.7in}

\section*{Appendix }

Let $F_{r,s, \delta}(\cdot)$ denote the noncentral $F$-distribution with degrees of freedom $r$ and $s$ and non-centrality parameter $\delta$, and let  $F_{r,s}(\cdot) = F_{r,s,0}(\cdot)$.  The mean and variance of  $F_{r,s, \delta}(\cdot)$ are
\begin{equation}
\frac{s(r+\delta)}{r(s-2)}  \ \text{ and } \
2 \, \frac{(r+\delta)^2 + (r+2\delta)(s-2)}
{(s-2)^2(s-4)},\label{dr02}
\end{equation}
assuming that $s > 2$ and $s>4$, respectively.

We use the following representation of these distributions (\citet{JKB}, eq.\ (30.10)),
\begin{eqnarray}
F_{r,s,\delta}(u)&=&\sum_{l=0}^{\infty}\frac{e^{-\frac{\delta}{2}}(\frac{\delta}{2})^l}{l!}F_{r+2l,s}\left(\frac{ru}{r+2l}\right)\label{nonF}\\
F_{r,s}(u)&=&I_{\frac{ru}{ru+s}}\left(\frac{r}{2},\frac{s}{2}\right)\label{F},
\end{eqnarray}
where $I_{u}(a,b)$ is the regularized incomplete beta function (i.e., beta distribution function) given by
\begin{eqnarray}\label{incompletebeta}
I_{u}(a,b)= \frac{1}{B(a,b)}\int_{0}^u t^{a-1}(1-t)^{b-1}dt,
\end{eqnarray}
$B(a,b)=\frac{\Gamma(a)\Gamma(b)}{\Gamma(a+b)}$ being the usual beta function.

\bigskip\noindent
{\bf Proof of Lemma~1.}
The conditional distribution of the projected data matrix ${\bf{X\mathbb{R}}}$ and ${\bf{Y\mathbb{R}}}$, given $\mathbb{R}$, are independent $N_k(\mathbb{R}'\mu_1, \mathbb{R}'\Sigma \mathbb{R})$ and $N_k(\mathbb{R}'\mu_2, \mathbb{R}'\Sigma \mathbb{R})$, respectively. Note that $S_\mathbb{R}=\mathbb{R}'S\mathbb{R}$, given $\mathbb{R}$, is distributed as Wishart $W_k\left(\frac1{n_1+n_2-1}\mathbb{R}'\Sigma \mathbb{R}, n_1+n_2-2\right)$.
According to Theorem 3.4.8 of \citet{Mardia},
\begin{eqnarray}\label{detS}
{\bf | }S_\mathbb{R}{\bf|} = {\bf |} \mathbb{R}'\Sigma \mathbb{R} {\bf |} \prod_{j=1}^k \chi_{n_1+n_2-j-1}^2,
\end{eqnarray}
where $\chi_{n_1+n_2-j-1}^2$ for $j=1,\ldots,k$ are independent $\chi^2$ random variables. From the expression (\ref{detS}), the proof is completed by showing that  $\lambda_{\min}\left(\mathbb{R}'\Sigma\mathbb{R}\right)>0$ with probability 1, where $\lambda_{\min}(A)$ is the minimum eigenvalue of the matrix $A$. Now, observe that
\begin{eqnarray*}
\lambda_{\min}\left(\mathbb{R}'\Sigma\mathbb{R}\right)&=&\inf_{||u||_2=1}u'\mathbb{R}'\Sigma\mathbb{R}u\\
&\ge& \inf_{||v||_2=1}v'\Sigma v \inf_{||u||_2=1} ||\mathbb{R}u||^2=\lambda_{\min}(\Sigma)>0.
\end{eqnarray*}
\hfill\(\Box\)

\bigskip\noindent
{\bf Proof of Theorem~1}

\noindent {\it Part (a).} Note that
\begin{equation}
E[\phi(T_\mathbb{R}^2)] = E_{{\mathbb{R}}}\left\{E_{_{{\bf X,Y}}}\left[\phi(T_\mathbb{R}^2)\big|\mathbb{R}\right]\right\}
= E_\mathbb{R}\left\{P_{_{{\bf X,Y}}}\left[\frac{n-k+1}{k}\cdot\frac{T^2_\mathbb{R}}{n}>c_\alpha\bigg| \mathbb{R}\right] \right\}.\label{dr03}
\end{equation}
Under ${\bf H_0}$, the conditional distribution of $\frac{n-k+1}{k}\ \frac{T^2_\mathbb{R}}{n}$ is $F_{k,n-k+1}$, independent of $\mathbb{R}$.   By \eqref{dr04}, we have
$
E[\phi(T_\mathbb{R}^2)|\bf{H_0}]= E_\mathbb{R}\left\{\alpha \right\}=\alpha.
$

\medskip

\noindent
{\it Part (b).}
Under ${\bf H_1^*}$ and for fixed $\mathbb{R}$, the conditional distribution of $\frac{n-k+1}{k}\, \frac{T^2_\mathbb{R}}{n}$ is  $F_{k,n-k+1,(n_1\inv+n_2\inv)\inv{\bf\Delta}_\mathbb{R}}$.
(Recall that
$
 {\bf \Delta}_\mathbb{R}=(\mu_1-\mu_2)'\mathbb{R}(\mathbb{R}'\Sigma \mathbb{R})^{-1}\mathbb{R}'(\mu_1-\mu_2).
$)
By \eqref{dr02} with $r= k$ , $s=n-k+1$, and $\delta=0$ we have that $c_\alpha \to 1$.   By \eqref{dr02} with $r= k$ , $s=n-k+1$, and $\delta = (n_1\inv+n_2\inv)\inv \Delta_\mathbb{R}$
we have  under $\bf{H}_1^*$, and for fixed $\mathbb{R}$, that the mean and variance $\frac{n-k+1}{k}\, \frac{T^2_\mathbb{R}}{n}$ behave asymptotically as
$c_\alpha + (n_1\inv+n_2\inv)\inv\Delta_\mathbb{R}/k$ and $2/n$, respectively. (We say that $a$ behaves asymptotically as $b$  if $a/b \to 1$.)

 It then follows from \eqref{dr01}, \eqref{dr04}, \eqref{dr03}, and Chebychev's inequality that
\begin{eqnarray}\label{powereq}
 E[\phi(T_\mathbb{R}^2)|\bf{H_1^*}] = E_{{\mathbb{R}}}\left\{E_{_{{\bf X,Y}}}\left[\phi(T_\mathbb{R}^2)\big|\mathbb{R},\bf{H^*_1}\right]\right\} \to 1.
\end{eqnarray}

\medskip

\noindent
{\it Part (c)}. By using the property that $I_{u}(a+1,b)\le I_{u}(a,b)$, and (\ref{F}), we have
\begin{eqnarray}\label{unbias}
I_{\frac{kc_\alpha}{kc_\alpha+n-k+1}}\left(\frac{k}{2}+l,\frac{n-k+1}{2}\right)&\le&I_{\frac{kc_\alpha}{kc_\alpha+n-k+1}}\left(\frac{k}{2},\frac{n-k+1}{2}\right)\nonumber\\
&=&F_{k,n-k+1}(c_\alpha)=1-\alpha.
\end{eqnarray}
Thus, by using (\ref{powereq}) and (\ref{unbias}), we have $E[\phi(T_\mathbb{R}^2)|\bf{H_1}]\ge\alpha$.
\hfill\(\Box\)

\bigskip\noindent
{\bf Proof of Theorem~2} By evaluating the conditional probability given the data matrix and subsequently taking expectation over that, we have
\begin{eqnarray}\label{Pr1}
P\left[\bar\theta^*<u\right]&=&E_{_{{\bf X,Y}}}\left\{P_{_{\mathbb{R}}}\left[\bar\theta^*<u\bigg|{\bf X,Y}\right]\right\}.
\end{eqnarray}
Note that
\begin{eqnarray}
P_{_{\mathbb{R}}}\left[\bar\theta^*<u\bigg|{\bf X,Y}\right]&=&P_{_{\mathbb{R}}}\left[\frac{\bar\theta^*-E_{_{\mathbb{R}}}\left(\theta_1^*\big|{\bf X,Y}\right)}{\sqrt{V_{_{\mathbb{R}}}\left(\theta_1^*\big|{\bf X,Y}\right)/m}} <\frac{u-E_{_{\mathbb{R}}}\left(\theta_1^*\big|{\bf X,Y}\right)}{\sqrt{V_{_{\mathbb{R}}}\left(\theta_1^*\big|{\bf X,Y}\right)/m}} \quad\bigg|{\bf X,Y}\right],\qquad
\end{eqnarray}
where $E_{_{\mathbb{R}}}\left(\theta_1^*\big|{\bf X,Y}\right)$ and $V_{_{\mathbb{R}}}\left(\theta_1^*\big|{\bf X,Y}\right)$ are conditional mean and variance of $\theta_1^*$ given the data matrix $\bf{X, Y}$. Further, given $\bf{X, Y}$, the random variables $\{\theta_i^*,~i=1,2\ldots,m\}$ are independent and identically distributed with finite variance. 
Now by using the Central Limit Theorem, we have
\begin{eqnarray}\label{cltm}
\lim_{m\rightarrow\infty}\left\{P_{_{\mathbb{R}}}\left[\bar\theta^*<u\bigg|{\bf X,Y}\right]-\Phi\left(\frac{u-E_{_{\mathbb{R}}}\left(\theta_1^*\big|{\bf X,Y}\right)}{\sqrt{V_{_{\mathbb{R}}}\left(\theta_1^*\big|{\bf X,Y}\right)/m}}\right)\right\}=0,\qquad
\end{eqnarray}
where $\Phi(\cdot)$ is the standard normal cumulative distribution function.
From (\ref{pval}),
\begin{eqnarray*}
&&\hskip-15ptE_{_{\mathbb{R}}}(\theta_1^*|{\bf X,Y})\notag\\&=&E_{_{\mathbb{R}}}\left[1-F_{ k,n-k+1}\left(\frac{n-k+1}{k}\cdot\frac{T_{\mathbb{R}_1}^2}{n}\right)\bigg|{\bf X,Y}\right]\notag\\
&=&\int \left\{1-F_{ k,n-\!k\!+\!1}\left(\!\!\frac{n-\!k\!+\!1}{k}\cdot\frac{\frac{n_1n_2}{n_1\!+\!n_2} (\overline{X}\!-\!\overline{Y})' R(R'SR)^{-1} R'(\overline{X}\!-\!\overline{Y})}{n_1\!+\!n_2\!-\!2}\!\right)\right\}d{\bf P}_{R},\qquad
\end{eqnarray*}
where ${\bf P}_{R}$ is the probability measure corresponding to random matrix $\mathbb{R}$. We claim that distribution of $E_{_{\mathbb{R}}}(\theta_1^*|{\bf X,Y})$ does not depend upon the parameters $\mu_1$, $\mu_2$ and $\Sigma$. To hold the claim, it suffices to show that
\begin{eqnarray}\label{claim1}
&&\hskip-30ptE_{{\bf X,Y}}\left[E_{_{\mathbb{R}}}(\theta_1^*|{\bf X,Y})\right]^r\notag\\
&=&
 \int\left[\!\int \left\{\!1\!-\!F_{ k,n-\!k\!+\!1}\left(\!\!\frac{n-\!k\!+\!1}{k}\cdot\frac{\frac{n_1n_2}{n_1\!+\!n_2} (\overline{X}\!-\!\overline{Y})' R(R'SR)^{-1} R'(\overline{X}\!-\!\overline{Y})}{n_1\!+\!n_2\!-\!2}\!\right)\!\right\}d{\bf P}_{R}\!\right]^r
d{\bf P}_{{\bf X,Y}}\notag\\
&&~\mbox{does not depend upon ($\mu_1,\mu_2,\Sigma$) for $r=1,2,\ldots$},
\end{eqnarray}
where ${\bf P}_{{\bf X,Y}}$ is the probability measure corresponding to the data matrix $\bf{X, Y}$.

Note that $0\le E_{_{\mathbb{R}}}(\theta_1^*|{\bf X,Y})\le 1$. Observe that
\begin{eqnarray}\label{cond_p}
&&\hskip-25pt\int\!\int \left\{\!1\!-\!F_{ k,n-\!k\!+\!1}\left(\!\!\frac{n-\!k\!+\!1}{k}\cdot\frac{\frac{n_1n_2}{n_1\!+\!n_2} (\overline{X}\!-\!\overline{Y})' R(R'SR)^{-1} R'(\overline{X}\!-\!\overline{Y})}{n_1\!+\!n_2\!-\!2}\!\right)\!\right\}^rd{\bf P}_{R}
d{\bf P}_{{\bf X,Y}}\notag\\
&\!\!\!=\!\!\!&\int\!\left[\int \left\{\!1\!-\!F_{ k,n-\!k\!+\!1}\left(\!\!\frac{n-\!k\!+\!1}{k}\cdot\frac{\frac{n_1n_2}{n_1\!+\!n_2} (\overline{X}\!-\!\overline{Y})' R(R'SR)^{-1} R'(\overline{X}\!-\!\overline{Y})}{n_1\!+\!n_2\!-\!2}\!\right)\!\right\}^r
d{\bf P}_{{\bf X,Y}} \right]\!d{\bf P}_{R},\quad
\end{eqnarray}
where interchange of integral are permitted by Fubini's theorem. Now, observe that under ${\bf H_0}$, the distribution of $F_{ k,n-\!k\!+\!1}\left(\!\!\frac{n-\!k\!+\!1}{k}\cdot\frac{\frac{n_1n_2}{n_1\!+\!n_2} (\overline{X}\!-\!\overline{Y})' R(R'SR)^{-1} R'(\overline{X}\!-\!\overline{Y})}{n_1\!+\!n_2\!-\!2}\!\right)$ is $U(0,1)$ for any given Projection matrix $R$. Therefore, the inner integral
\begin{eqnarray}\label{inner_int}
&&\int \left\{\!1\!-\!F_{ k,n-\!k\!+\!1}\left(\!\!\frac{n-\!k\!+\!1}{k}\cdot\frac{\frac{n_1n_2}{n_1\!+\!n_2} (\overline{X}\!-\!\overline{Y})' R(R'SR)^{-1} R'(\overline{X}\!-\!\overline{Y})}{n_1\!+\!n_2\!-\!2}\!\right)\!\right\}^r
d{\bf P}_{{\bf X,Y}}\notag\\&&~\mbox{does not depend upon the parameter $(\mu_1,\mu_2,\Sigma)$}.
\end{eqnarray}
This imply that (\ref{cond_p}) does not depend upon the parameter for any positive integer $r$.

Now note that, from (\ref{claim1}) and by using Fubini theorem, we have
\begin{eqnarray}\label{final_step}
&&\hskip-20ptE_{{\bf X,Y}}\left[E_{_{\mathbb{R}}}(\theta_1^*|{\bf X,Y})\right]^r\notag\\
&&\hskip-20pt=
 \int\ldots\int\left[\!\int \prod_{i=1}^r\left\{\!1\!-\!F_{ k,n-\!k\!+\!1}\left(\!\!\frac{n-\!k\!+\!1}{k}\cdot\frac{\frac{n_1n_2}{n_1\!+\!n_2} (\overline{X}\!-\!\overline{Y})' R_i(R_i'SR_i)^{-1} R_i'(\overline{X}\!-\!\overline{Y})}{n_1\!+\!n_2\!-\!2}\!\right)\!\right\}\!d{\bf P}_{{\bf X,Y}}\right]
\prod_{i=1}^r d{\bf P}_{R_i}\notag\\
&&
\end{eqnarray}
We can view that $R_i$ for $i=1,\ldots,r$ are iid with probability measure $P_{R}$ in the expression (\ref{final_step}). By using this and (\ref{inner_int}), it follows that
$$\int \prod_{i=1}^r\left\{\!1\!-\!F_{ k,n-\!k\!+\!1}\left(\!\!\frac{n-\!k\!+\!1}{k}\cdot\frac{\frac{n_1n_2}{n_1\!+\!n_2} (\overline{X}\!-\!\overline{Y})' R_i(R_i'SR_i)^{-1} R_i'(\overline{X}\!-\!\overline{Y})}{n_1\!+\!n_2\!-\!2}\!\right)\!\right\}
d{\bf P}_{{\bf X,Y}},$$
does not depend upon the parameter $(\mu_1,\mu_2,\Sigma)$ which in turn imply that (\ref{claim1}) holds for any positive integer $r$.
Similarly, under ${\bf H_0}$, the distribution of $V_{_{\mathbb{R}}}\left(\theta_1^*\big|{\bf X,Y}\right)$ too does not depend on the parameters.
Now note that
\begin{eqnarray}\label{cltm1}
\left|P_{_{\mathbb{R}}}\left[\bar\theta^*<u\bigg|{\bf X,Y}\right]-\Phi\left(\frac{u-E_{_{\mathbb{R}}}\left(\theta_1^*\big|{\bf X,Y}\right)}{\sqrt{V_{_{\mathbb{R}}}\left(\theta_1^*\big|{\bf X,Y}\right)/m}}\right)\right|<2.\qquad
\end{eqnarray}
From (\ref{Pr1}), (\ref{cltm}), (\ref{cltm1}) and the dominated convergence theorem, we have
\begin{eqnarray*}
\lim_{m\rightarrow\infty}\left\{P\left[\bar\theta^*<u\right]
-E_{{\bf X, Y}}\left[\Phi\left(\frac{u-E_{_{\mathbb{R}}}\left(\theta_1^*\big|{\bf X,Y}\right)}{\sqrt{V_{_{\mathbb{R}}}\left(\theta_1^*\big|{\bf X,Y}\right)/m}}\right)\right]\right\}=0
\end{eqnarray*}
Thus, for any $n_1, n_2$, as $m\rightarrow\infty$, the asymptotic distribution of $\frac1m\sum_{i=1}^m\theta_i^*$ does not depend on the parameters $\mu_1,\mu_2$, and $\Sigma$. This completes the proof.
\hfill\(\Box\)

\bigskip\noindent
{\bf Proof of Theorem~3} The power of the test (\ref{randomemptest}) is
\begin{eqnarray*}
E[\phi^*|{\bf H_1^*}]&=& P\left[\bar\theta^*<u_{\{\alpha,n_1,n_2\}}\bigg|{\bf H_1^*}\right],
\end{eqnarray*}
where $u_{\{\alpha,n_1,n_2\}}$ is such that $$ P\left[\bar\theta^*<u_{\{\alpha,n_1,n_2\}}\bigg|{\bf H_0}\right]=\alpha.$$
For a given $\alpha$, $n_1$, and $n_2$, we have $0<u_{\{\alpha,n_1,n_2\}}<1$. Thus, there exists a convergent subsequence of  $u_{\{\alpha,n_1,n_2\}}$. With an abuse of the notation, let this subsequence be $u_{\{\alpha,n_1,n_2\}}$, converging to $u_\alpha$.

We claim that $u_\alpha >0$.  To see this, note first that for all $(n_1, n_2)$, $P(\bar\theta^* \le \epsilon |{\bf H_0}) \le P(m^{-1} \theta_1 \le \epsilon | {\bf H_0})  = \epsilon m$, since $\theta_i$ is uniform(0,1) distributed under $\bf H_0$.
Thus, there exist positive $\epsilon$ such that $P(\bar\theta^* \le \epsilon | {\bf H_0}) <  \alpha$ for all $(n_1,n_2)$.  It follows that $u_{\alpha,n_1,n_2} \ge \epsilon $ for all $(n_1,n_2)$ and therefore $u_\alpha \ge \epsilon > 0$.

Let $\nu$ be positive.
Since $\theta_i$ is the p-value of the test $\phi(T^2_\mathbb{R})$, it follows from Theorem 1 (b) with $\alpha = \nu$ that
$P(\theta_i < \nu | {\bf H_1^*}) = P(\phi(T^2_\mathbb{R})=1| {\bf H_1^*}) \to 1$. Therefore, since $m$ is fixed and finite, $P(\theta_i < \nu, \ i=1,\ldots,m | {\bf H_1^*}) \to 1$ and consequently,
$P(\bar\theta^* < \nu | {\bf H_1^*}) \to 1$ .  This result holds for all $\nu >0$.
Since $u_{\{\alpha,n_1,n_2\}} \to u_\alpha >0$, it follows that $P(\bar\theta^* < u_{\{\alpha,n_1,n_2\}} | {\bf H_1^*}) \to 1$, that is, $\lim_{n_1,n_2\rightarrow\infty}E[\phi^*|{\bf H_1^*}]= 1$.

\hfill\(\Box\)

\section*{Acknowledgement}

Radhendushka Srivastava was a postdoctoral researcher supported by NSF-DMS 0808864 and NSF-EAGER 1249316. The work of Ping Li is  supported by ONR-N000141310261, NSF-III-1360971,  NSF-BIGDATA-1419210, and AFOSR-FA9550-13-1-0137.


\begin{thebibliography}{}

\bibitem[Alon et~al., 1999]{Alon}
Alon, U., Barkai, N., Notterman, D.~A., Gish, K., Ybarra, S., Mack, D., and
  Levine, J. (1999).
\newblock Broad patterns of gene expression revealed by clustering analysis of
  tumor and normal colon tissues probed by oligonucleotide arrays.
\newblock {\em Proc. Natl. Ecad. Sci. USA}, 96:6745--6750.

\bibitem[Bai and Sarandasa, 1996]{BS}
Bai, Z. and Sarandasa, H. (1996).
\newblock Effect of high dimension: By an example of a two sample problem.
\newblock {\em Statistica Sinica}, 6:311--329.

\bibitem[Benjamini and Hochberg, 1995]{BH}
Benjamini, Y. and Hochberg, Y. (1995).
\newblock Controlling the false discovery rate: A practical and powerful
  approach to multiple testing.
\newblock {\em J. Roy. Statist. Soc. Ser. B}, 57:289--300.

\bibitem[Charikar et~al., 2004]{Article:Charikar_2004}
Charikar, M., Chen, K., and Farach-Colton, M. (2004).
\newblock Finding frequent items in data streams.
\newblock {\em Theor. Comput. Sci.}, 312(1):3--15.

\bibitem[Chen et~al., 2011]{CPPW}
Chen, L.~S., Paul, D., Prentice, R.~L., and Wang, P. (2011).
\newblock A regularized hoteeling's \text{T2} test for pathway analysis in
  proteomic studies.
\newblock {\em J. Amer. Statist. Assoc.}, 106(496):1345--1360.

\bibitem[Chen and Qin, 2010]{CQ}
Chen, S.~X. and Qin, Y.~L. (2010).
\newblock A two-sample test for high-dimensional data with application to
  gene-set testing.
\newblock {\em Ann. Statist.}, 38:808--835.

\bibitem[Chen et~al., 2010]{CZZ}
Chen, S.~X., Zhang, L.~X., and Zhong, P.~S. (2010).
\newblock Tests for high-dimensional covariance matrices.
\newblock {\em J. Amer. Statist. Assoc.}, 105:810--819.

\bibitem[Clemencon et~al., 2009]{9}
Clemencon, S., Depecker, M., and Vayatis, N. (2009).
\newblock \text{AUC} optimization and the two-sample problem.
\newblock {\em Advances in Neural Information Processing Systems}.

\bibitem[Cuesta-Albertos et~al., 2007]{11}
Cuesta-Albertos, J.~A., Barrio, E.~D., Fraiman, R., and Matran, C. (2007).
\newblock The random projection method in goodness of ﬁt for functional data.
\newblock {\em Computational Statistics and Data Analysis}, 51(10):4814--4831.

\bibitem[Diaconis and Freedman, 1984]{15}
Diaconis, P. and Freedman, D. (1984).
\newblock Asymptotics of graphical projection pursuit.
\newblock {\em Annals of Statistics}, 12(3):793--815.

\bibitem[Fan et~al., 2007]{FHY}
Fan, J., Hall, P., and Yao, Q. (2007).
\newblock To how many simultaneous hypothesis tests can normal, student's t or
  bootstrap calibration be applied.
\newblock {\em J. Amer. Statist. Assoc.}, 102:1282--1288.

\bibitem[Goeman and Buhlmann, 2007]{2}
Goeman, J.~J. and Buhlmann, P. (2007).
\newblock Analyzing gene expression data in terms of gene sets: methodological
  issues.
\newblock {\em Bioinformatics}, 23(8):980--987.

\bibitem[Jacob et~al., 2010]{10}
Jacob, L., Neuvial, P., and Dudoit, S. (2010).
\newblock Gains in power from structured two-sample tests of means on graphs.
\newblock {\em Technical Report: arXiv:q-bio/1009.5173v1}.

\bibitem[Johnson et~al., 1995]{JKB}
Johnson, N.~L., Kotz, S., and Balakrishnaha, N. (1995).
\newblock {\em Continuous Univariate Distributions}, volume~2.
\newblock Wiley, New York, 2nd edition.

\bibitem[Kosorok and Ma, 2007]{KM}
Kosorok, M. and Ma, S. (2007).
\newblock Marginal asymptotics for the ``large p, small n'' paradigm: With
  applications to microarray data.
\newblock {\em Ann. Statist.}, 35:1456--1486.

\bibitem[Kuelbs and Vidyashankar, 2010]{KV}
Kuelbs, J. and Vidyashankar, A. (2010).
\newblock Asymptotic inference for high-dimensional data.
\newblock {\em Ann. Statist.}, 38:836--869.

\bibitem[Ledoit and Wolf, 2002]{LW}
Ledoit, O. and Wolf, M. (2002).
\newblock Some hypothesis tests for the covariance matrix when the dimension is
  large compared to the sample size.
\newblock {\em Ann. Statist.}, 30:1081--1102.

\bibitem[Li and Chen, 2012]{LC}
Li, J. and Chen, S.~X. (2012).
\newblock Two sample tests for high-dimensional covariance matrices.
\newblock {\em Ann. Statist.}, 40:908--940.

\bibitem[Li et~al., 2006]{Li06}
Li, P., Hastie, T.~J., and Church, K.~W. (2006).
\newblock Very sparse random projections.
\newblock {\em Proceedings of the 12th ACM SIGKDD international conference on
  Knowledge discovery and data mining}, pages 287--296.

\bibitem[Li et~al., 2011]{Li11}
Li, P., Shrivastava, A., Moore, J.~L., and Konig, A.~C. (2011).
\newblock Hashing algorithms for large-scale learning.
\newblock {\em Proceedings of the 24th Annual Conference on Neural Information
  Processing Systems (NIPS)}, pages 2672--2680.

\bibitem[Lopes et~al., 2012]{LJW}
Lopes, M.~E., Jacob, L.~J., and Wainwright, M.~J. (2012).
\newblock A more powerful two-sample test in high dimension using random
  projection.
\newblock {\em arXiv:1108.2401v2 [math.ST]}.

\bibitem[Lu et~al., 2005]{1}
Lu, Y., Liu, P., Xiao, P., and Deng, H. (2005).
\newblock Hotelling’s \text{T2} multivariate profiling for detecting
  differential expression in microarrays.
\newblock {\em Bioinformatics}, 21(14):3105–--3113.

\bibitem[MacKinnon, 2009]{MacKinnon}
MacKinnon, J.~G. (2009).
\newblock {\em Bootstrap hypothesis testing (Handbook of Computational
  Econometrics (Edited by D. A. Belsley and E. Kontoghiorghes))}.
\newblock John Wiley and Sons, West Sussex.

\bibitem[Mardia et~al., 1979]{Mardia}
Mardia, K.~V., Kent, J.~T., and Bibby, J.~M. (1979).
\newblock {\em Multivariate Analysis}.
\newblock Academic Press Inc., London.

\bibitem[Marzetta et~al., 2011]{12}
Marzetta, T.~L., Tucci, G.~H., and Simon, S.~H. (2011).
\newblock A random matrix–theoretic approach to handling singular covariance
  estimates.
\newblock {\em IEEE Transactions on Information Theory}, 57(9):6256--6271.

\bibitem[Srivastava, 2007]{Sri}
Srivastava, M.~S. (2007).
\newblock Multivariate theory for analyzing high dimensional data.
\newblock {\em J. Japan Statist. Soc.}, 37:53--86.

\bibitem[Srivastava and Du, 2008]{SD}
Srivastava, M.~S. and Du, M. (2008).
\newblock A test for the mean vector with fewer observations than the
  dimension.
\newblock {\em J. Multivariate Anal.}, 99:386--402.

\bibitem[van~der Laan and Bryan, 2001]{VB}
van~der Laan, M. and Bryan, J. (2001).
\newblock Gene expression analysis with the parametric bootstrap.
\newblock {\em Biostatistics}, 2:445--461.

\bibitem[Vempala, 2004]{14}
Vempala, S.~S. (2004).
\newblock The random projection method.
\newblock {\em DIMACS Series in Discrete Mathematics and Theoretical Computer
  Science, American Mathematical Society}.

\bibitem[Ville et~al., 2004]{3}
Ville, D. V.~D., Blue, T., and Unser, M. (2004).
\newblock Integrated wavelet processing and spatial statistical testing of fmri
  data.
\newblock {\em Neuroimage}, 23(4):1472--1485.

\bibitem[Wang et~al., 2013]{WPQ}
Wang, R., Peng, L., and Qi, Y. (2013).
\newblock Jackknife empirical likelihood test for equality of two high
  dimensional means.
\newblock {\em Statistica Sinica}, page dx.doi.org/10.5705/ss.2011.261.

\end{thebibliography}
\end{document}